\documentclass{article}

\usepackage{iclr2021_conference,times}

\usepackage{graphicx}
\usepackage{cite}
\usepackage{hyperref}
\usepackage{soul}
\usepackage{color}
\usepackage{multirow}
\usepackage{tabularx}

\usepackage{tikz}

\title{Attacks, defenses, and tools: a framework to facilitate robust AI/ML systems}

\author{Mohamad Fazelnia, Igor Khokhlov, Mehdi Mirakhorli \\
Rochester Institute of Technology\\
Rochester, NY 14623, USA \\
\texttt{\{mf8754,ixk8996,mxmvse\}@rit.edu}
}

\iclrfinalcopy 
\begin{document}


\maketitle

\begin{abstract}
Software systems are increasingly relying on Artificial Intelligence (AI) and Machine Learning (ML) components. The emerging popularity of AI techniques in various application domains attracts malicious actors and adversaries. Therefore, the developers of AI-enabled software systems need to take into account various novel cyber-attacks and vulnerabilities that these systems may be susceptible to. This paper presents a framework to characterize attacks and weaknesses associated with AI-enabled systems and provide mitigation techniques and defense strategies. This framework aims to support software designers in taking proactive measures in developing AI-enabled software, understanding the attack surface of such systems, and developing products that are resilient to various emerging attacks associated with ML. The developed framework covers a broad spectrum of attacks, mitigation techniques, and defensive and offensive tools. In this paper, we demonstrate the framework architecture and its major components, describe their attributes, and discuss the long-term goals of this research.

\end{abstract}



\section{Introduction} \label{sec:introduction}
Today's software systems increasingly rely on various artificial intelligence (AI) and machine learning (ML) components. For instance, flight control software of commercial airlines rely on AI-enabled autopilot components, medical diagnostic software products rely on AI for image analysis, autonomous driving vehicles heavily rely on object detection and recognition, and modern intrusion detection systems rely on various ML-based classification models and anomaly detection.
This increase in AI and ML (AI/ML) deployment in software products can be explained by various reasons, such as availability of the data needed for AI/ML models training, more powerful computers capable of much faster model training, drastic advances in the AI/ML algorithms, etc. 

Followed by this popularity of AI/ML-based systems, attack and defense strategies on AI/ML systems have been widely studied \citep{szegedy2014intriguing,goodfellow2015explaining, 7780651, 7546524}. Also, several works have done surveys on this domain \citep{8290925, xu2020adversarial, pitropakis2019taxonomy, yuan2019adversarial} and tried to categorize attacks and defenses based on their attributes and assumptions.
However, we still lack a comprehensive analysis of the techniques and tactics used in this domain.

During software design and development, designers have to understand all potential threats related to the AI/ML algorithms deployment and leverage mitigation techniques in order to address them. While there have been extensive resources for weaknesses~\citep{CWE} and attacks~\citep{attck} on software systems, there are not many organized knowledge bases for attacks, weaknesses, and mitigation techniques for AI-enabled systems.

To this end, we present an \href{http://design.se.rit.edu/programs/ai-ml-framework}{\textcolor{blue}{online framework\footnote{\href{http://design.se.rit.edu/programs/ai-ml-framework}{www.design.se.rit.edu/programs/ai-ml-framework}}}} that aims to enumerate possible attacks to the AI-enabled systems, identifies appropriate mitigation techniques, tools and toolchains used by adversaries, and characterizes attackers. This framework provides a structured and scalable comprehensive knowledge base that can help researchers, practitioners, and AI capability builders be better informed of attacks, toolchains, and mitigation techniques and guide them to perform attack modeling of their AI-enabled software systems and reason about adversaries, identify them, and respond intelligently.

While this paper presents the framework's architecture, the long-term goal is to develop a framework that enables cyber-risk analysis of AI-enabled Software systems. To the best of our knowledge, this is the first work that provides a comprehensive framework based on a systematic literature review that categorizes adversarial actors and explains their behavior, techniques and tactics used by them, their appropriate mitigation techniques, and the offensive and defensive tools used in AI/ML systems. The proposed project is a fundamental step toward developing a knowledge base for various offensive and defensive measurements, representations, and other mechanisms for securing AI-enabled software systems.

\section{Methodology} \label{sec:methodology}

This work aims to provide a comprehensive framework that contains all the attacks, mitigation techniques, and tools in AI/ML domain. To this end, we conducted a systematic literature review of articles published in the two of the most popular digital libraries, IEEE Xplore \footnote{\href{https://ieeexplore.ieee.org/Xplore/home.jsp}{www.ieeexplore.ieee.org}} and ACM Digital Library. \footnote{\href{https://dl.acm.org}{www.dl.acm.org}} We explored all peer-reviewed papers published between 2000 and 2020 years. The first step includes searching in papers' abstract using such keywords (and their variations) as ``artificial intelligence,'' ``machine learning,'' ``deep learning,'' ``neural networks,'' ``mitigation,'' ``threat,'' and ``adversarial.'' 

As the initial result, we gathered around 14500 papers in total. Then, two experts in the area filtered out irrelevant papers based on their titles, which resulted in 2500 filtered papers.
The next step included paper filtration based on their abstracts. Next, we performed a detailed literature review. 
During reviewing each paper, we also employed a snowballing process to review other related works mentioned in the literature that have not been published in the two aforementioned digital libraries.
After we analyzed the first random 10\% of the selected papers, we were able to develop a primary skeleton of the proposed framework. 
Based on this review, we identified the attributes that could characterize different attacks, as well as the techniques and tactics they use to violate the systems. 
 
In order to build the proposed framework, we perform the following steps: 
\begin{enumerate}
\itemsep0em
    \item Conduct a systematic literature review to gather information representing attacks on AI/ML algorithms, mitigation strategies, and tools related to attacks or mitigation techniques. 
    \item Analyze and identify emerging attacks on AI/ML algorithms.
    \item Classify attacks, tools, and mitigation strategies from various perspectives and develop appropriate components of the framework. This step resulted in a set of attributes for each component of the framework, which are presented in Section \ref{sec:architecture}.
    \item Develop a bi-directional mapping between all components and implement the developed framework in a form that can be used by security practitioners and system developers. 
    \item Finally, the framework has to be validated by gathering feedback from system designers that used our framework, which is out of this paper's scope.
\end{enumerate}

This paper briefly discusses the first two steps and focuses on the third step, and briefly touches on the fourth step. The fifth step will be presented in future work.
\section{Framework to Describe AI/ML ATT\&CK} \label{sec:architecture}

\begin{figure}[htp]
    \centering
    \includegraphics[width=\textwidth]{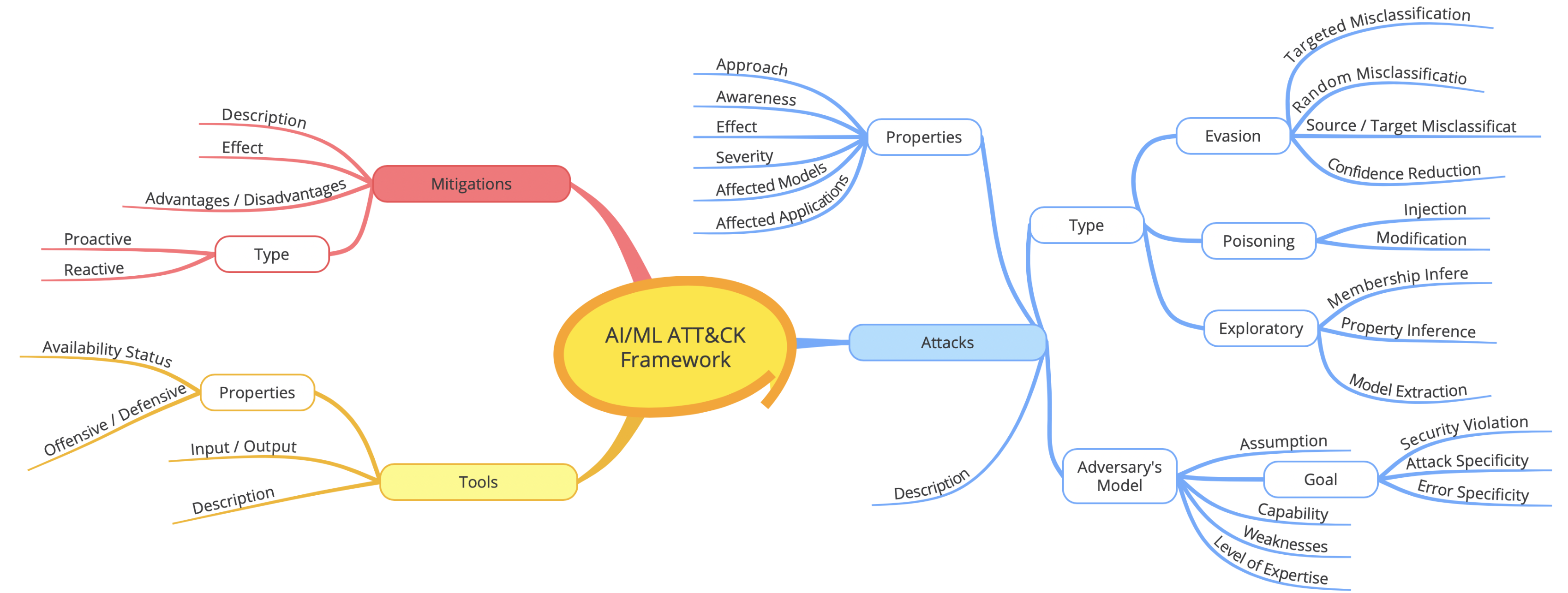}
    \caption{An overview of the meta model}
    \label{fig:mindmap}
\end{figure}

Based on the conducted systematic literature review and the analysis of the selected papers, we developed the framework with three major components: \emph{attacks}, \emph{mitigation techniques}, and \emph{tools}, and a set of attributes for each component.
Each component provides a comprehensive review of its area. The framework's mindmap is presented in figure \ref{fig:mindmap}.

\subsection{Attacks}

This section expands the \emph{attacks} component and describes its characteristics briefly, as well as its relationship to the \emph{mitigation} and \emph{tools} components.

\subsubsection{Types}
Attacks on AI/ML-enabled systems are categorized into three types, \emph{Poisoning attacks, Exploratory attacks, and Evasion attacks}. In \textbf{Poisoning Attacks} the attacker tries to undermine the learning process during model training via two approaches, \emph{injection} and \emph{modification}, in which the attacker injects malicious samples into the training process \citep{10.1145/3128572.3140451} and the attacker directly changes the training data  \citep{biggio2011support}, respectively. \textbf{Exploratory Attacks} aim to extract private information from the system, which can be categorized to the following groups: \emph{Membership inference}, in which the attacker aims to determine whether a particular data exists in the training dataset \citep{7958568}. In \emph{property inference} the attacker tries to extract information about the model and the training data, like data distribution, or the model hyper-parameters. Finally, in \emph{model extraction}, the attacker tries steal the ML model and architecture \citep{197128}.
\textbf{Evasion Attacks} refer to those attacks that try to create a malicious sample that the model classifies mistakenly. The attacker could reduce the confidence level of the predicted output which is called (\emph{confidence reduction}), or based on the input and target class, the attacks are categorized to \emph{misclassification}, \emph{target misclassification} and \emph{source-target misclassification} \citep{7467366}.

\subsubsection{Attributes}
To better characterize the attacks on AI/ML systems, understanding the attack's attributes is extremely important. 
To this end, we considered the following attributes for attacks:
\textbf{Approach} explains the mechanism of the attack.
\textbf{Effect} shows the consequences of the attack.
\textbf{Awareness} illustrates the level of information that is available to the attacker.
\emph{White-box} \citep{carlini2017evaluating} denotes full access to the model's architecture, data and the parameters, while in a \emph{black-box} setting, neither data nor the model is available to the attacker.
The \emph{gray-box} setting refers to the attacks with limited knowledge about the target model and its data \citep{9149635}.
\textbf{Severity} indicates the level of the violation.
\textbf{Mitigation} presents appropriate mitigation strategies and maps the attack to the related part of the ``mitigation'' component of the framework.
\textbf{Affected Models} shows the models that are vulnerable to the attack. 
\textbf{Affected Applications} mentions some of the datasets, systems, and real-world applications that are susceptible to the attack.



\subsubsection{Adversary's Model}    
In addition to the attack attributes, we include adversary model to capture a threat agent that has the power to, exploit a vulnerability or conduct other damaging activities in AI-enabled software.
To model the adversary, the following characteristics are considered:
\textbf{Goal} expresses the specific attack goal and  motivation. \textbf{Security Violation} illustrates the type of violation, which could be one or a combination of these types: \emph{integrity violation}, in which the adversary undermines the normal system's behavior, in \emph{availability violation}, the attacker compromises the normal performance of the model, and in \emph{privacy violation} the attacker obtains private information about the model or the data \citep{10.1145/3128572.3140451}.
\textbf{Attack specificity} shows if the attacker aims to affect only a specific subset of the samples, which is called \emph{targeted}, or any sample in the system, which is called \emph{indiscriminate}.
\textbf{Error specificity} shows whether the attacker wants to misclassify the samples into the desired class or any classes other than the true class, which is called \emph{error specific} and \emph{error generic}, respectively \citep{10.1145/3128572.3140451}.
\textbf{Assumption} explains the adversary's access over the model and its resources. \textbf{Capabilities} indicates the abilities and the malicious potentials of the attack. \textbf{Weaknesses} represents the adversary's weaknesses and its blind-spots. Finally,
\textbf{Level of Expertise} shows the required skill level to perform the attack.

\subsection{Mitigation Techniques}
The ``mitigation'' component presents available techniques and approaches that can be used to mitigate or defend from the attacks.
Each entry in this component is mapped to the entries of the ``attacks'' component and has the following attributes:
\textbf{Approach} briefly describes the mitigation method. 
\textbf{Effect} explains how utilizing the mitigation technique makes the system more robust against attacks.
\textbf{Type} indicates if the technique is defending the model before the attack happens (\emph{proactive}) \citep{7546524}, or the technique is responding to the attack after the attack happened (\emph{reactive}) \citep{grosse2017statistical}. 
\textbf{Advantage \& Disadvantage} assesses the benefits and drawbacks of utilizing the mitigation technique, considering its efficiency, cost, and side effects.

\subsection{Tools}
This component provides information about tools that enable users to deploy attacks on AI/ML algorithms, assess the model robustness, and provide mitigation techniques to make their model more robust against the attacks.
Similar to the two previous components, entries of this component are mapped to entries of ``attacks'' and ``mitigation'' components. For each tool, the following attributes are considered: 
\textbf{Input} and \textbf{Output} indicate the type and format of the inputs and outputs of the tool.
\textbf{Offensive/Defensive} indicates whether a tool is called to carry out an attack or is designed to defend against the attacks.
\textbf{Availability Status} describes the tool's availability, whether it is available to the public, open-source, commercial, revised from a prior toolchain, or it is entirely new.


\subsection{Implementation}
We collect the information we obtained from our systematic literature review in online knowledge base captured as tables. The tables' rows correspond to the techniques discussed in the paper, and the columns represent different attributes discussed earlier in this section. Then we used this information to develop the  \href{http://design.se.rit.edu/programs/ai-ml-framework}{\textcolor{blue}{online framework\footnote{\href{http://design.se.rit.edu/programs/ai-ml-framework}{www.design.se.rit.edu/programs/ai-ml-framework}}}}. 
The framework is provided with several filters, which allows users to focus on a specific problem and get into details as necessitated, evaluate the problem from different viewpoints, and find an appropriate solution with respect to the advantages, disadvantages, and side effects of the solution.
\section{Conclusion} \label{sec:conclusion}
In this work, we presented a framework to establish a  comprehensive knowledge base of attacks on AI/ML systems with appropriate mitigation techniques and defensive and offensive tools.
To develop the presented meta model of the framework, we performed a systematic literature review and analyzed selected papers from various perspectives.
The presented meta model (framework's skeleton) thoroughly describes all information required to know about various attacks, mitigation techniques, and tools and describes the relationship between them.
The paper presents initial attributes for each component, which will be extended in the immediate future.
Even though the presented framework is in its initial phase, it is capable of covering a broad spectrum of attacks on AI/ML algorithms in various applications and domains.





\bibliography{references.bib}
\bibliographystyle{iclr2021_conference}


\end{document}